# A Temporal Tree Decomposition for Generating Temporal Graphs

Corey Pennycuff
University of Notre Dame
Notre Dame, Indiana 46556
cpennycu@nd.edu

Salvador Aguinaga
University of Notre Dame
Notre Dame, Indiana 46556
saguinag@nd.edu

Tim Weninger
University of Notre Dame
Notre Dame, Indiana 46556
tweninge@nd.edu

## ABSTRACT

Discovering the underlying structures present in large real world graphs is a fundamental scientific problem. Recent work at the intersection of formal language theory and graph theory has found that a Hyperedge Replacement Grammar (HRG) can be extracted from a tree decomposition of any graph. This HRG can be used to generate new graphs that share properties that are similar to the original graph. Because the extracted HRG is directly dependent on the shape and contents of the of tree decomposition, it is unlikely that informative graph-processes are actually being captured with the extraction algorithm. To address this problem, the current work presents a new extraction algorithm called temporal HRG (tHRG) that learns HRG production rules from a temporal tree decomposition of the graph. We observe problems with the assumptions that are made in a temporal HRG model. In experiments on large real world networks, we show and provide reasoning as to why tHRG does not perform as well as HRG and other graph generators.

## 1 INTRODUCTION

The discovery and analysis of network patterns is central to the scientific enterprise. Thus, extracting the useful and interesting building blocks of a network is critical to the advancement of many scientific fields. Indeed the most pivotal moments in the development of a scientific field are centered on discoveries about the structure of some phenomena [12]. For example, chemists have found that many chemical interactions are the result of the underlying structural properties of interactions between elements [3].

Biologists have agreed that tree structures are useful when organizing the evolutionary history of life [5, 9], sociologists find that triadic closure underlies community development [7], and neuroscientists have found "small world" dynamics within the brain [2]. Unfortunately, current graph mining research deals with small predefined patterns [10] or frequently reoccurring patterns [4, 8, 13], even though interesting and useful information may be hidden in unknown and non-frequent patterns.

Pertinent to this task are algorithms that learn the LEGO-like building blocks of real world networks in order to gain insights into the mechanisms that underlie network growth and evolution. In pursuit of these goals, Aguinaga *et al.* recently discovered a relationship between graph theory and formal language theory that allows for a *Hyperedge Replacement Grammar* (HRG) to be extracted from the *tree decomposition* of any graph [1]. Like a context free grammar (CFG), but for graphs, the extracted HRG contains the precise building blocks of the network as well as the instructions by which these building blocks ought to be pieced together. In addition, this framework is able to extract patterns of the network's structure from small samples of the graph in order to generate networks that have properties that match those of the original graph.

In their typical use-case, CFGs are used to represent and generate patterns of strings through rewriting rules. A natural language parser, for example, learns how sentences are recursively built from smaller phrases and individual words. In this case, it is important to note that the CFG production rules used by natural language parsers encode the way in which sentences are logically constructed, that is, the CFG contains descriptive information about how nouns and verbs work together to build coherent sentences. CFGs can therefore generate new sentences that are at least grammatically correct. This is not the case with HRGs.

On the contrary, the HRG is completely dependant on the graph's tree decomposition, otherwise known as the clique tree, junction tree, or cluster tree, depending on the context. Unfortunately, there are many ways to perform a tree decomposition on a given graph, and even the optimal, *i.e.*, minimal-width, tree decomposition is not unique. As a result, the production rules in a standard HRG are unlikely to represent the informative structural rules of a graph. The problem is clear: in order to deeply understand the growth and evolution of a graph we need the tree decomposition to not only represent the structures within a graph, but also encode how these structures grow and evolve over time.

In the present work we address this problem through a temporal HRG extraction algorithm (tHRG) that maintains the important graph generation guarantees and properties of the original HRG algorithm, while also creating production rules that encode how the graph evolves over time. This is accomplished by preserving a graph's temporal changes within the structure of the tree decomposition, so that the production rules of the extracted HRG contain descriptive information about how the graph evolves. This solution also permits rule-history to be computed from the tree decomposition.

The present work shows negative results. In initial experiments on 4 temporal networks, we show that the tHRG model makes improper assumptions and is unable to model the growth of a temporal graph. The development of the tHRG model was a significant engineering effort, so we present these negative so that others need not duplicate our efforts.

## 1.1 Tree Decomposition

Before we describe our method, some background definitions are needed. We begin with an arbitrary input *hypergraph* $H = (V, E)$, where a *hyperedge* $e \in E$ can connect multiple vertices. Common *graphs* (*e.g.*, social networks, Web graphs, information networks) are a particular case of hypergraphs where each edge connects exactly two vertices.

For convenience, all of the graphs in this paper will be *simple*, *connected* and *undirected*, although these restrictions are not vital. In the remainder of this section we refer mainly to previous developments in tree decomposition and their relationship to hyperedge replacement grammars.

All graphs can be decomposed (though not uniquely) into a *tree decomposition*. Within the data mining community, tree decompositions are best known for their role in exact inference in probabilistic graphical models [11].

## 1.2 Hyperedge Replacement Grammars

HRGs are a graphical counterpart to CFGs used in compilers and natural language processing [6]. Like in a CFG, an HRG contains a set of production rules $\mathcal{P}$, each of which contains a left hand side (LHS) $A$ and a right hand side (RHS) $R$. In context free string grammars, the LHS must be a nonterminal character, which can be replaced by some set of nonterminal or terminal characters on the RHS of the rule. In HRGs, nonterminals are graph-cliques and a RHS can be any graph (or hypergraph) fragment.

Just as a CFG generates a string, an HRG can generate a graph by repeatedly choosing a nonterminal $A$ and rewriting it using a production rule $A \rightarrow R$. The replacement hypergraph fragment $R$ can itself have other nonterminal hyperedges, so this process is repeated until there are no more nonterminals in the graph.

Tree decompositions and HRGs have been studied separately for some time in discrete mathematics and graph theory literature. HRGs are conventionally used to generate graphs with very specific structures, *e.g.*, rings, trees, stars. A drawback of many current applications of HRGs is that their production rules must be manually defined. For example, the production rules that generate a ring-graph are distinct from those that generate a tree, and defining even simple grammars by hand is difficult or impossible. Very recently, Kemp and Tenenbaum developed an inference algorithm that learned probabilities of the HRG's production rules from real world graphs, but they still relied on a handful of rather basic hand-drawn production rules (of a related formalism called vertex replacement grammar) to which probabilities were learned [10]. Kukluk, Holder and Cook were able to define a grammar from frequent subgraphs [4, 8, 13–15], but their methods have a coarse resolution because *frequent* subgraphs only account for a small portion of the overall graph topology.

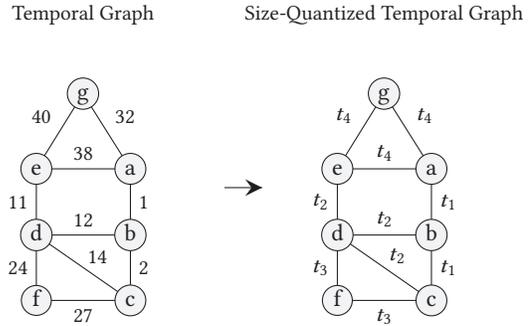

**Figure 1: Temporal graph showing edge creation times (at left). Edge creation times are quantized into 4 equal-sized bins (with arbitrary tiebreaking) and the timestamps are reassigned to their bin number (on right).**

In earlier work we showed that an HRG can be extracted from a static graph and used to generate new graphs that maintained the same global and local properties as the original graph. We also proved under certain (impractical) circumstances that the HRG can be used to generate an isomorphic copy of the original graph [1].

Prior work in HRGs extract their production rules directly from a tree decomposition, but because the tree decomposition can vary significantly, the production rules may also vary significantly. Although good at generating new synthetic graphs that are similar to the original graph, the production rules of HRGs do not describe the growth process that created the graph. In this section we show how to represent growth rules from *temporal graphs* in a tree decomposition. The temporal HRG (tHRG) extracted from the temporal tree decomposition will therefore contain production rules that describe the growth of the graph.

The tHRGs framework is divided into two phases: 1) temporal tree decomposition, *i.e.*, the temporal rule extraction phrase, and 2) a graph generation phase, where the extracted rules are applied to generate a new graph.

## 2 TEMPORAL HYPEREDGE REPLACEMENT GRAMMARS

We represent temporal graphs as an edgelist, where an edge $e^{(i)}$ represents the interaction of a set of nodes at some particular time. We iterate through this edgelist in order from earliest time $t^{(0)}$ to latest time $t^{(n)}$, adding edges and vertices as needed. As in most graph datasets, we assume that once an edge or vertex is created it is never deleted; under this assumption, temporal graphs are sometimes referred to as *cumulative graphs*.

### 2.1 Size Quantization

In most graph datasets, edges are added one at a time with a timestamp with millisecond precision. Although it may be important to understand fine-grained edge creation in some cases, in the present work we coarsen the time precision in order to resolve the evolution of more complex structures rather than a single edge at a time. We create $\beta$ edge-bins and reassign each edge's time to its bin number.

We considered two types of binning approaches: equally sized bins or equally timed bins. In the present work we use size-quantization because it breaks up the computational effort into equal-sized chunks. Figure 1 illustrates a temporal graph (at left) and as a size-quantized temporal graph where $\beta=4$ (on right).

## 2.2 Graph De-Evolution

The focus of the present work is to extract an HRG with production rules that represent the structural events that occur within a growing graph. Because the HRG directly depends on a tree decomposition of a graph, it is necessary to first create a tree decomposition based on the the structural changes that are found in the growth of the graph. This section presents an algorithm that creates a temporal tree decomposition for the purposes of extracting a tHRG, with an example to follow.

---

**Algorithm 1** Constructing Temporal HRGs

1: **function** TEMPORALTREEDECOMPOSITION($H = \{V, E\}$)
2:     $N, T \leftarrow []$
3:     **for** $e^{(i)} \leftarrow t_n$ **to** $t_0$ **do**
4:         $H^{(i)} \leftarrow$ INDUCEDSUBGRAPH($H, e^{(i)}$)
5:         **while** $C^{(i)} \leftarrow$ MAXCLIQUE($H^{(i)}$) **do**    // [17]
6:             $T$.ADDNODE($\eta^{(i)} \leftarrow$ ASTERMINAL($C^{(i)}$))
7:             $H^{(i)} \leftarrow H^{(i)} \setminus C_x^{(i)}$
8:             $H \leftarrow H \setminus C_x^{(i)}$
9:             $v \leftarrow V_{\eta^{(i)}} \setminus E$    // orphans
10:            **if** $v \neq \{\emptyset\}$ **then**
11:               **for each** $\eta$ in $T$ s.t. $v \in \eta \wedge \eta$ not marked **do**
12:                   $\eta' \leftarrow$ MERGE($\eta, \eta'$)
13:               **end for**
14:               $\eta' \leftarrow \eta' \setminus v$    // external nodes of $\eta^{(i)}$
15:               **if** $\eta' = \{\emptyset\}$ **then**
16:                   $\eta' \leftarrow$ S    // S is the starting nonterminal
17:               **end if**
18:               $T$.ADDNODE($\eta' \leftarrow$ ASNONTERMINAL($V_{\eta'}$))
19:               $V = V \setminus v$
20:            **else**
21:               $T$.ADDNODE($\eta' \leftarrow$ ASNONTERMINAL($V_{\eta^{(i)}}$))
22:            **end if**
23:             $T$.ADDEDGE($\eta' \rightarrow \eta^{(i)}$)
24:            MARK($\eta^{(i)}$)
25:         **end while**
26:     **end for**
27:     **return** $T$
28: **end function**

---

We build the tree decomposition from the bottom-up using a process that reverses the growth of the graph, and extracts the tree decomposition along the way. Pseudo-code for this process is shown in Alg. 1. We take, as input, any graph or hypergraph $H = \{V, E\}$, and initialize an empty tree decomposition $T$ and set of non-terminals $N$. The process begins by selecting only those edges $e^{(i)}$ that appeared in timestep $i$, starting from the most recent time $t_n$. With those edges we find the induced subgraph from $H$ called $H^{(i)}$, which represents the graphical structures that were added in $i^{th}$ timestep. We iteratively extract the largest clique from $H^{(i)}$ labeled $C_x^{(i)}$ [17]. The edges in the maximal clique are labeled as terminal edges and a tree decomposition node $\eta_x^{(i)}$ is added to $T$. The edges in $C_x^{(i)}$ are then removed from $H$ and $H^{(i)}$ in lines 7 and 8 respectively.

Recall that our temporal graph representation relied specifically on the creation of edges. A side effect of that representation is that nodes can only appear when they are first connected to the graph. Therefore, the edges added in the current timestamp, and consequently deleted in lines 7 and 8, may represent the addition of a node, which needs to be accounted for in the model. Because of the edge removal, those nodes $v \in C_x^{(i)}$ may now be orphaned $v \notin E$, i.e., they are no longer incident to any edges. In the case where no nodes are orphaned by the removal of edges, we add a tuple $\langle$AsNONTERMINAL($V_{\eta_x^{(i)}}$), $\eta_x^{(i)} \rangle$ consisting of a new nonterminal edge containing the vertices in $\eta_x^{(i)}$ and the $\eta_x^{(i)}$-node itself in line 27. In this case, $\eta_x^{(i)}$ can only contain terminal edges and must therefore be a leaf node in $T$.

In the case where the removal of edges creates orphaned nodes, we iterate through all of the orphans in line 11. Although these orphan nodes are no longer incident to any edges in $E$, i.e., terminal edges, they may still be incident to several nonterminal edges in $N$. In order to ensure that the running intersection property of tree decompositions is maintained, it is important to encode these instances in $T$. Fortunately, by virtue of the cumulative, i.e., growth-only, assumption of our graphs, the now orphaned node $v$ cannot contain any connections from a previous timestep. This allows us to include the non-terminals incident to the orphan $v$ into its tree-node $\eta_x^{(i)}$ in line 15. Because $N$ stores a tuple of the incident nonterminal and its referenced tree-node $\eta'$, we must add an edge between $\eta_x^{(i)}$ and $\eta'$ in $T$.

To create the parent nonterminal of $\eta_x^{(i)}$ we need to find its *external vertices u* by removing the orphaned vertex (aka the inner vertices) from $\eta_x^{(i)}$. If there are not external vertices, then $\eta_x^{(i)}$ has no parent, so we use the special starting nonterminal S instead. Each tree decomposition can only have one starting nonterminal and it will always be the at the root of the tree. We add the external vertices as a nonterminal edge to $N$ along with a pointer to $\eta_x^{(i)}$ for future reference. Finally we remove the nonterminal edges extracted in this iteration from $N$ and remove the orphaned vertex from the graph.

## 2.3 Example Temporal Tree Decomposition

For an intuitive example of how this algorithm works, we will demonstrate Alg. 1 using the example graph illustrated in Fig 1. This graph has 7 nodes (a-g) and 10 edges across 4 time-quantized bins; this example graph was chosen because it exhibits all of the necessary boundary conditions needed to fully describe the extraction algorithm.

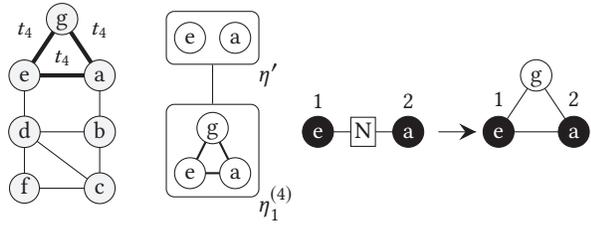

Figure 2: Example of hyperedge replacement grammar rule creation from an interior vertex of the tree decomposition in timestamp 4. Note that lowercase letters inside vertices are for explanatory purposes only; only the numeric labels outside external vertices are actually part of the rule.

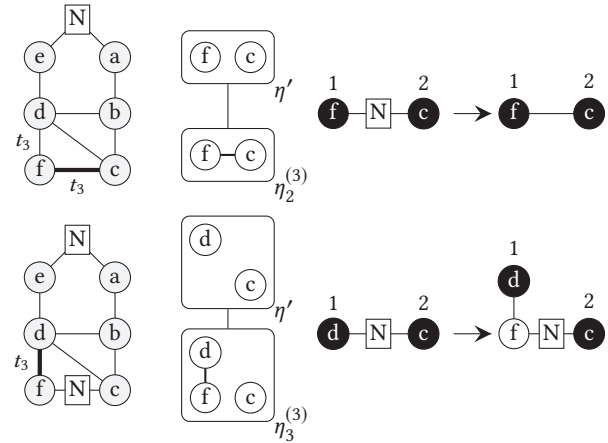

Figure 3: Example grammar creation of timestep 3

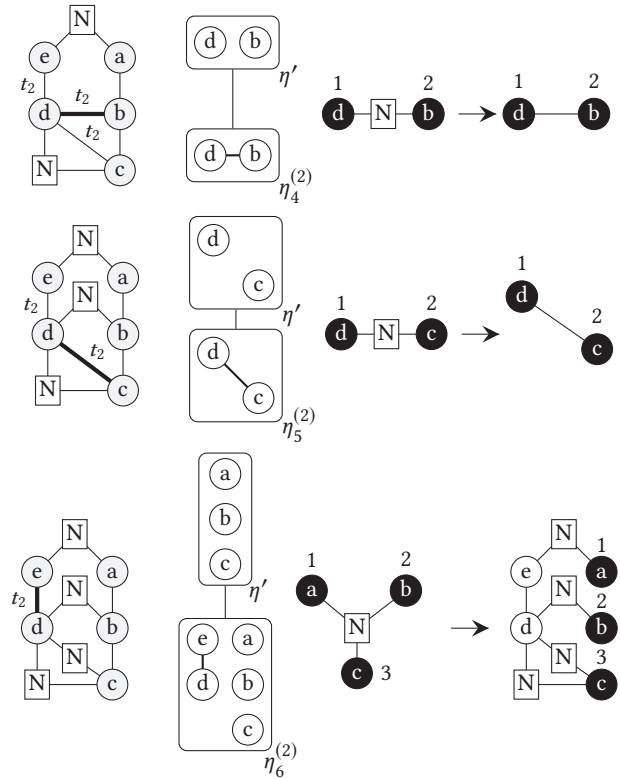

Figure 4: Example grammar creation of timestep 2

Beginning with the edges added in the latest timestep, $t_4$, we induce a subgraph (line 4) and find its maximal clique, *i.e.*, the three node triangle $C_x^{(i)}$=(a,e,g) (line 5), and add it as a nonterminal ⟨a,e,g⟩ to $N$ (line 6). We also add the edges of the (a,e,g)-triangle as terminal-edges in a new tree decomposition node, $\eta_1^{(4)}$, which is added to $T$ (line 7). Removing the (a,e,g)-triangle's edges from the graph (lines 8 and 9) reveals a new orphan-vertex (g) (line 7). The orphaned node, $v$, is incident only to the original nonterminal $C = C_1^{(4)}$ (line 14), so no new nonterminals are added to $\eta_1^{(4)}$. The orphaned node (g) is subtracted from the (a,e,g) leaving only (a,e) as the external nodes $u$ (line 17). The external nodes form a nonterminal tuple ⟨a,e⟩ and a new node $\eta'$ that is added to $T$ (line 21) as the parent of the (a,e,g)-triangle (line 22). Finally, the ⟨a,e⟩ nonterminal is added to the set of nonterminals (line 23), the nonterminal-clique ⟨a,e,g⟩ is removed from $N$ (line 24), and the orphaned node is removed from the graph (line 25).

Figure 2 illustrates the iteration through timestamp $t_4$ with the tree decomposition linking the new $\eta'$ containing the external ⟨a,e⟩ nodes with the (a,e,g)-triangle in $\eta_1^{(4)}$. From the edge that is created between $\eta'$ and any $\eta_x^{(i)}$ an HRG production rule can be extracted using the method described by Aguinaga et al [1]. For this example the production rule is illustrated in the right-half of Fig 2.

The next-latest edge is from timestamp $t_3$. As illustrated in Fig. 3, we induce a subgraph and select the maximal clique. In this case there are two equally maximal cliques: the edges (d,f) and (c,f). We chose the (c,f)-edge at random, add it as a terminal edge to a new tree decomposition node $\eta_2^{(3)}$, and remove it from $H$ and $H^{(i)}$. The removal of (c,f) does not reveal any orphans, so we simply add a new nonterminal ⟨c,f⟩ as $\eta'$ to $T$ as the parent of $\eta_2^{(3)}$.

Next we choose (d,f) because it is the only remaining clique from $t_3$. Like before we add the clique edges as terminal edges to $T$ as a new node $\eta_3^{(3)}$. The removal of (d,f) from the graph results in an orphan-node f, which already exists in the unmarked parent of $\eta_2^{(3)}$. So we merge these tree decomposition nodes, remove f from the graph and add a nonterminal between external nodes d and c.

This process continues likewise for the timestep 2. Then finishes with timestep 1, which contains two cliques (illustrated on the following page).

This example will create the tree decomposition found in Fig. 6. There are several critical observations that can be made from this tree decomposition. First, this result is very different from an optimal or near-optimal tree decomposition of the same static graph. This observation is important and expected because we hope to learn HRG rules from the growth of the graph, rather than learning from its static representation.

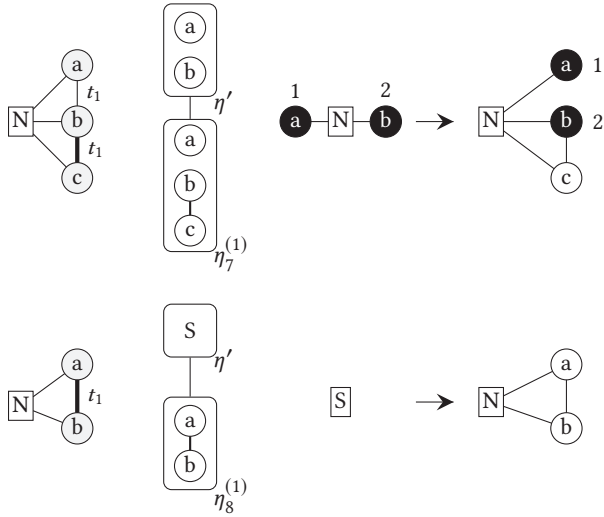

Figure 5: Example grammar creation of timestep 1

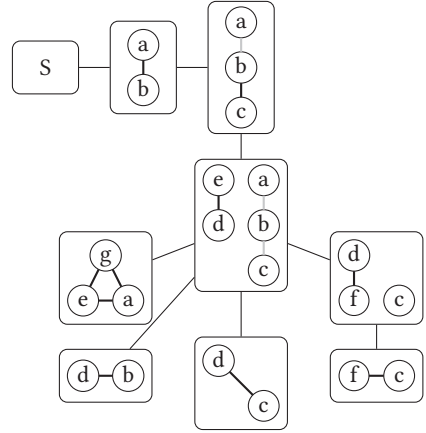

Figure 6: Example of hyperedge replacement grammar rule creation from an interior vertex of the tree decomposition. Note that lowercase letters inside vertices are for illustrative purposes only.

Unfortunately, this graph does not look like how we expect a tree decomposition to look like in general. The width of the tree is 4 (*i.e.*, the size of largest tree-node |(a,b,c,e,d)| - 1 = 4), which is twice the width of the optimal static tree decomposition. Finally, although Fig. 6 is a mathematically proper tree decomposition (*i.e.*, its HRG will reconstruct an isomorphic copy of the original graph), it does not "look" like a the tree decompositions that we are accustomed to seeing. For example, the largest nodes are almost always seen at the top of the tree, rather than in the middle and towards the leaves.

### 2.4 Problems Caused by Late Bridges

After lengthy study, we found that the reason for the oddness of the trees created by our temporal tree decomposition algorithm is due to the the existence of "late bridges" that occur in many temporal graphs. In the conventional way of thinking, late bridges are those edges that connect disparate nodes late in the growth of the graph. For example, a late bridge in a social network may be a pair of new co-workers that have become friends through work, but are otherwise highly separated in the network.

Observe that the depth of the tree decomposition node corresponds to the time from which the node was extracted. Late-stage connections (*e.g.*, timestamp 3 and 4) can be found at the bottom of the tree decomposition, and early-stage connections (*e.g.*, timestamp 1 and 2) can be found towards the top of the tree decomposition. Because of the running intersection property and because our temporal tree decomposition algorithm works backwards in time, we have to carry late bridge connections through the entire tree decomposition until their link is resolved – like tracing how distant relatives are related in a family tree. As a result, the width of the tree decomposition increases significantly with every late bridge.

Static graphs are immune to this problem because they are free to optimize the tree decomposition unconstrained by the edge creation time. Nevertheless, the grammars created by the tree decomposition may still represent the growth patterns of the graph. To test this hypothesis, we will use the temporal grammars to generate graphs and compare the generated graphs with the original.

### 2.5 Growing Graphs with tHRG

Here we describe how to use the tHRG to generate graphs. In larger HRGs we usually find many production rules that are identical. Rather than storing all rule-instances, we merge duplicates and keep a count of the number of times that each distinct rule has been seen. Duplicate merging creates a probabilistic HRG (pHRG) where rule probabilities correspond to the number of times an rule has been seen.

To generate a random graph $H'$ from a probabilistic HRG, we start with the special starting nonterminal $H' = S$. From this point, a new graph $H^*$ can be generated as follows: (1) Pick any nonterminal $A$ in $H'$; (2) Find the set of rules associated with LHS $A$; (3) Randomly choose one of these rules with probability proportional to its count; (4) replace $A$ in $H'$ with the rule's RHS to create $H^*$; (5) Replace $H'$ with $H^*$ and repeat until there are no more nonterminal edges.

We find that although the generated graphs have the same mean size as the original graph, the variance is much too high to be useful. So we want to sample only graphs whose size is the same as the original graph's, or some other user-specified size. Naively, we can do this using rejection sampling: sample a graph, and if the size is not right, reject the sample and try again. However, this would be quite slow. Our implementation uses a dynamic programming approach to do this exactly while using quadratic time and linear space, or approximately while using linear time and space. We omit the details of this algorithm here, but the source code is available online at https://github.com/nddsg/PHRG/.

### 2.6 Problems Caused by Chomsky Normal Form

The probabilistic algorithm used to generate exact-sized graphs requires that the tHRG is in Chomsky Normal Form (CNF). CNF

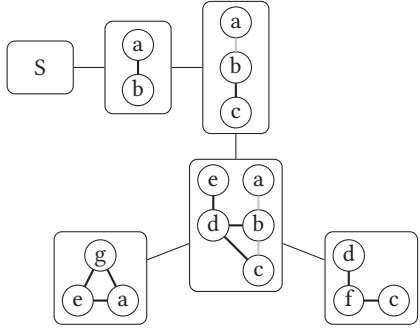

**Figure 7: Chomsky Normal Form of the example graph.**

requires 1) that each rule's right-hand side has at most two non-terminals and 2) that each rule's right-hand side has at least one internal vertex (*i.e.*, a vertex in a node $\eta$ that does not appear in its parent).

**Pruning.** To ensure the tree decomposition is in CNF we use the following scheme. Let $\eta$ be a tree decomposition node with children $\eta_1, \ldots, \eta_d$ and a parent $\eta'$. If $\eta$ does not contain any internal vertices, we remove $\eta$ from the tree decomposition and rewire its children (if any) to be children of $\eta'$.

**Binarization.** By the HRG induction methods presented in pervious work [1], the number of children of $\eta$ determines the number of nonterminals on the right-hand side of the resulting rule. Thus, to be in CNF it suffices for the tree decomposition to have a branching factor of at most two. Although the branching factor of a tree decomposition may be greater than two, it is always easy to binarize it: For $\eta$ where $d > 2$ (here $d$ corresponds to the number of children for a given parent node). Make a copy of $\eta$; call it $\eta^a$, and let $V_{\eta^a} = V_\eta$. Let the children of $\eta$ be $\eta_1$ and $\eta^a$, and let the children of $\eta^a$ be $\eta_2, \ldots, \eta_d$. Then, if $\eta^a$ has more than two children, apply this procedure recursively to $\eta^a$.

Unfortunately, due to the early bridges problem the largest nodes in the tree decomposition are also the most likely to need binarized. Each time we copy $\eta$ in the binarization step we exacerbate the early bridges problem and further complicate the model. To continue the example from Fig. 6, a binarized tree decomposition is illustrated in Fig. 7.

Figure 7 shows a the example tree decomposition from Fig. 6 that removes three tree decomposition nodes because they did not have any internal vertices. No binarization steps were needed. As a result of the CNF pruning, the rules created by $\eta_2^{(3)}$, $\eta_4^{(2)}$, and $\eta_5^{(3)}$ are removed and the rules in $\eta_3^{(3)}$ and $\eta_6^{(2)}$ are updated to reflect the changes.

## 3 EXPERIMENTS

We use a methodology similar to the static HRG experiments. Given a (temporal) graph we generate a (temporal) tree decomposition and extract temporal HRG rules (tHRG). Beginning with the special starting nonterminal we apply the HRG grammar rules to generate a graph of size *n*, and compare the generated graph with the original graph across several local and global graph metrics. We also compare our graph generation results to temporal Exponential Random Graph Models (tERGM), as well as the standard ERGM, the Kronecker graph model, the Chung-Lu graph model, and the Block Two-level Erdős-Rényi (BTER) model by treating the final version of a temporal graph as a static graph.

### 3.1 Datasets

We begin with four small, publicly available network datasets. The exact time of all node/edge arrivals is known in all network datasets. Basic statistics for our four networks are shown in Table 1. Each dataset represents interactions between people. The Haggle dataset indicates when covered cellphone users contact each other. The Infectious and Hypertext datasets show near-field-communication interactions between conference badges at the 2009 INFECTIOUS Conference and the 2009 HyperText Conference. The Manufacturing Company dataset are from email-traces from 2010 of a mid-sized manufacturing company.

**Table 1: Network Dataset**

| Dataset Name | Type | Nodes | Edges | T(days) |
|---|---|---|---|---|
| Haggle | Cell Phone | 274 | 2,124 | 4 |
| Infectious | Conference | 410 | 2,765 | 0.33 |
| Hypertext | Conference | 113 | 2,196 | 3 |
| Mfg. Company | Email | 167 | 5,784 | 272 |

We consider each graph to be a *cumulative network*, where a timestamped interaction adds an edge to an ever-growing network. This is an important assumption because if an interaction between two individuals occurs more than once, then only the first interaction is considered. Therefore, our graphs are temporal, cumulative, undirected, and simple.

### 3.2 Methodology and Results

The general goal of graph generation is to create synthetic graphs that maintain several of the same properties as the original graph. In the case of temporal graphs our goal is to learn how to generate graphs by modelling the growth process directly.

HRG rules can hold a history of their ancestry from the tree decomposition by including grandparent-nodes, great-grandparent-nodes, etc. on the LHS of the production rule. If needed, during the graph generation process, a tHRG rule can be matched to a hyperedge in a growing graph as well as its history. Although adding history to the model may increase model accuracy, it will also increase the model complexity. For example, keeping a full ancestry of each HRG rule is the same as keeping the entire tree decomposition, which is as big as the original graph. Yet, by keeping the full tree decomposition we are guaranteed an isomorphic copy of the original graph. As a tradeoff between space complexity and modelling accuracy we introduce a history parameter $\alpha$, where $\alpha = 0$ means no history is kept except the LHS and $\alpha = 1$ requires that one level of history be kept so that an application of an HRG rule must match the current LHS and as well as the LHS's parent in the tree decomposition.

As in Fig. 1, the first step is to create bins from the raw timestamps. There is no clear way to do this, so we introduced a binning parameter that creates equal sized bins of size $\beta$. For example, when $\beta$ = 100 the first 100 interactions are grouped into the first bin, the second set of 100 non-redundant interactions are grouped into the second bin, and so on.

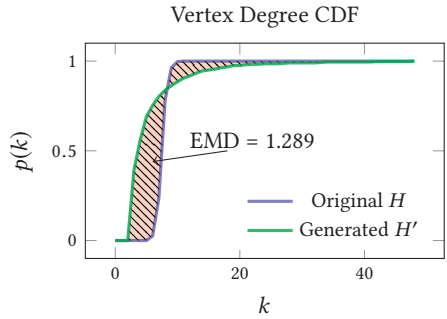

**Figure 8: Example comparison of the cumulative degree distribution of two graphs. The Earth Mover Distance (EMD) is the cumulative differences between the two distributions (highlighted in red).**

Besides the creation of a synthetic graph, the tHRG method also generates a hyperedge replacement grammar that the user may inspect to understand how the of the graph evolved over time. We are working on a way to evaluate the human interpretability of the temporal graph grammar, but for now we will compare the results of the final state of the original graph, with the graphs generated by tHRG as well as the state of the art graph generators introduced above.

There are dozens of ways to compare the generated graphs with the accumulated final-state of the original temporal graph. Graphs generated using tHRG, pHRG, Kronecker, Chung-Lu, BTER and tERGM are analyzed by studying their fundamental network properties to assess how successful the model performs in generating graphs from parameters and production rules learned from the input graph. We look at degree distributions and the hopplot to draw conclusions on these results.

Visually inspecting the results of these tests is challenging. We observe that each plot is essentially a distribution of values. So rather than visually inspecting the results of each possible distribution we compare the cumulative distribution function (CDF) of each metric with the CDF from the original graph.

Figure 8 shows an example cumulative degree distribution for the original Hypertext graph and a randomly generated Watts-Strogatz graph fitted to the original graph. Visual inspection clearly shows that these distributions are dissimilar, where their differences are highlighted in red. We adapt the Earth Mover Distance (EMD) to measure the area between these discrete CDFs. To avoid illustrating all pairwise combinations, we compare various graphs metrics using the EMD test instead. Lower is better.

For example, Fig. 9 shows how the EMD changes as $\beta$ increases from 50 to 500. Marks are the mean-rules of 50 trials from each graph generator. The 95% confidence interval is also shown, but the error bars too small to see in most cases. Recall that pHRG, Kronecker, Chung-Lu and Erdos-Reyni graphs are static graphs and therefore do not use the $\beta$ binning parameter so any variation from a flat line can be attributed to random noise in the averages.

The Chung-Lu method, which directly models the degree distribution of a graph, is the winner in the degree tests in Fig. 9,

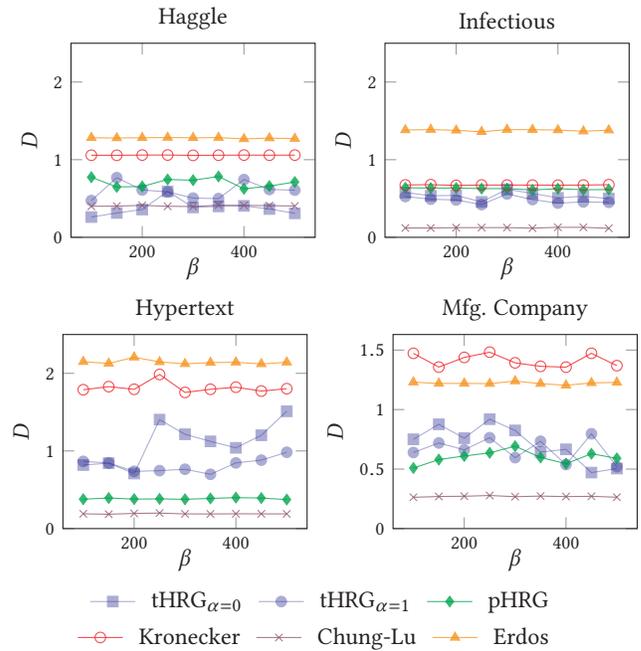

**Figure 9: EMD and confidence intervals for degree distributions (lower is better)**

followed closely by the static pHRG. Both variants of tHRG have a reasonable performance, which tends to worsen as $\beta$ increases. Keeping a rule-history (i.e., when $\alpha$=1) tends to perform better, but not enough to justify a model size that is twice as large as the the no-history variant.

We attempted to fit ERGM and tERGM using the ERGM R-package, but the models failed to converge.

Figure 10 shows a similar EMD on the hop plot distribution of generated graphs. Again, we attribute any variation in the results from static graph generators to randomness. Therefore, we interpret the results in Fig. 10 to be inconclusive. That is, each model creates graphs with hop plots that are equally-similar to the original graph. We performed similar tests on other global graph metrics such as the sorted eigenvector centrality and the sorted local clustering coefficient. We found similar inconclusive results.

Rather than looking at the degree distribution and the hop plot, there is mounting evidence which argues that the graphlet comparisons are the most complete way measure the similarity between two graphs [16, 18]. The graphlet distribution succinctly describes the number of small, local substructures that compose the overall graph and therefore more completely represents the details of what a graph "looks like." Furthermore, it is possible for two very dissimilar graphs to have the same degree distributions, hop plots, etc., but it is difficult for two dissimilar graphs to fool a comparison with the graphlet distribution. We therefore employ the graphlet correlation distance (GCD), which measures the frequency of the various graphlets present in each graph, i.e., the number of edges, wedges, triangles, squares, 4-cliques, etc., and compares the graphlet frequencies between two graphs. Because the GCD is a distance metric, lower values are better. The GCD can range from [0, +∞], where the GCD is 0 if the two graphs are isomorphic.

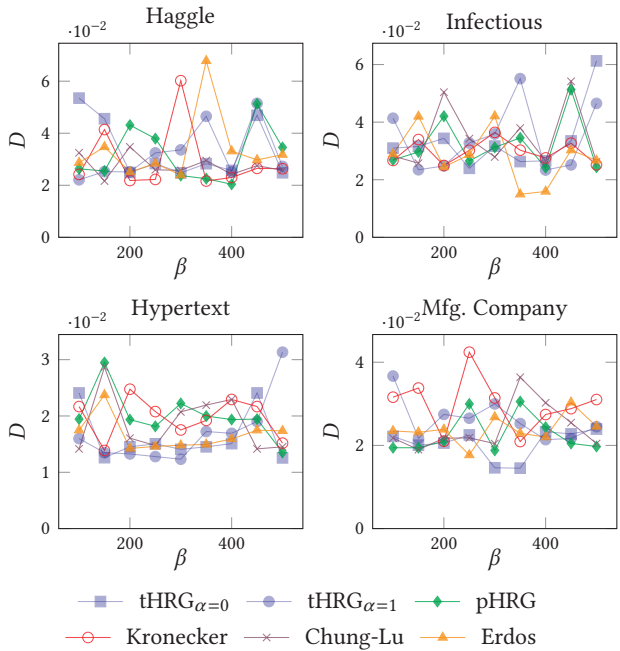

**Figure 10: EMD and confidence intervals for hopplot distributions (lower is better)**

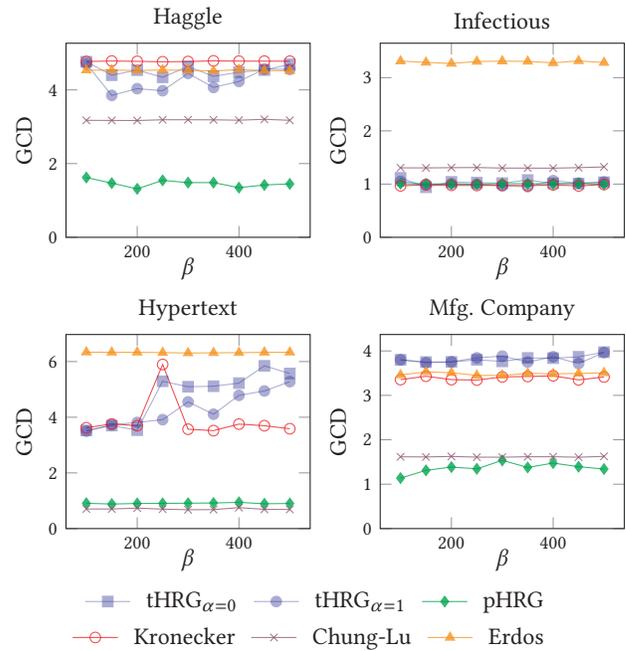

**Figure 11: Mean and confidence intervals for GCD (lower is better)**

The GCD is a single value, not a distribution. So we plot the mean and 95% confidence interval in Fig. 11. We find that the static pHRG performs the best or close to the best across the dataset. The tHRG model performs poorly on the GCD metric indicating that the tHRG method does not create graphs that "look like" the original graph.

## 4 CONCLUSIONS

In the present work we expand on previous work in probabilistic hyperedge replacement grammars (pHRG) to create a temporal adaptation that learns rules from the growth of the temporal graph. Our temporal adaptation is able to create a temporal tree decomposition that can be translated into a temporal graph grammar. This temporal grammar has the ability to generate an isomorphic copy of the original graph if the rules are applied in the correct (temporal) order. If we throw out the ordering and instead apply rules probabilistically, then we find that the generated graphs are **not** similar to the original graph. We believe that the reason for these negative results is because the temporal tree decomposition is constructed in an unnatural way. The result is a suboptimal tree decomposition caused by edges that act as early bridges in the graph. The existence of early bridges combined with the running intersection property results in tree decompositions with large widths and convoluted grammar rules even on the small graphs tested here.